\documentclass[aps,prb,twocolumn,superscriptaddress,longbibliography]{revtex4-1}

\usepackage{graphicx}
\usepackage{bm}
\usepackage{hyperref}
\usepackage{url}
\usepackage{natbib}
\usepackage{epstopdf}
\usepackage{epsfig}
\usepackage{hyperref}
\usepackage[none]{hyphenat}
\usepackage{xcolor}
\usepackage{caption,subcaption}
\usepackage{mathtools}
\usepackage{marvosym,wasysym}
\usepackage{relsize}
{ \hypersetup{
		colorlinks=true,
		linkcolor=blue,
		citecolor = blue,      
		urlcolor=blue,
	}         
	\newcommand{\RNum}[1]{\uppercase\expandafter{\romannumeral #1\relax}}
	\makeatother
\begin{document}
		
		\title{Entropic Stabilization and Descriptors of Structural Transformation in High Entropy Alloys}
		
		\author{Narendra Kumar}
		\affiliation{Chemistry and Physics of Materials Unit, Jawaharlal Nehru Centre for Advanced Scientific Research (JNCASR), Bangalore, India - 560064}
		\affiliation{Theoretical Sciences Unit, Jawaharlal Nehru Centre for Advanced Scientific Research (JNCASR), Bangalore, India - 560064}
		\author{Umesh V. Waghmare}
		\email{waghmare@jncasr.ac.in}
		\affiliation{Theoretical Sciences Unit, Jawaharlal Nehru Centre for Advanced Scientific Research (JNCASR), Bangalore, India - 560064}
		
		
		\begin{abstract}
			With first-principles theoretical analysis of the local structure using Bond Orientational Order parameters and Voronoi partitioning, we establish (a) HCP$\rightarrow$BCC structural transformation in high-entropy alloys (HEAs) Nb$_x$(HfZrTi)$_y$ at 16\% Nb-concentration, and (b) that the internal lattice distortions (ILDs) peak at the transition. We demonstrate that the relative stability of HCP and BCC structures is driven by energetics, while the overall stability is achieved with contribution from the vibrational entropy that exceeds the configurational entropy of mixing. We show that along with atomic size mismatch, low average number ($< 5$) of valence electrons and disparity in the crystal structures of constituent elements are responsible for larger ILDs in Nb$_x$(HfZrTi)$_y$ than in HEAs like Nb$_a$Mo$_b$W$_c$Ta$_d$. 
		\end{abstract}
		

	\maketitle
\section{Introduction}Alloying has been central to the progress of human civilization since the Bronze age. In conventional alloys, small amounts of secondary elements are mixed with primary ones. The resulting alloy is named based on the primary element like ferrous, aluminum, copper, and nickel alloys. The last two decades~\cite{yeh2004nanostructured,cantor2004microstructural,zhang2014,miracle2017critical,grosse2018data,george2019high,murty2019high} have witnessed an unconventional method of alloy design with equiatomic mixing of four or more elements. Stability of such alloy is assumed to be dominated by configurational entropic contributions, and hence they are named high-entropy alloys (HEAs). Beginning with the work of Yeh et al.~\cite{yeh2004nanostructured} and Cantor et al.~\cite{cantor2004microstructural}, HEAs have stimulated intense research to develop understanding of their phase stability and superb mechanical behavior~\cite{zhang2014,miracle2017critical,grosse2018data,george2019high,li2019fcc}.
\par
Four \emph{core effects}~\cite{pickering2016high} govern the stability and behavior of HEAs: 1) high configurational entropy of mixing attributed to stabilizing solid solution phase, 2) severe lattice distortions due to mismatch in the chemistry of alloying elements, 3) sluggish diffusion kinetics, and 4) the cocktail effect resulting in extraordinary properties. While the enhanced configurational entropy lowers the Gibbs free energy, it is not the sole factor responsible for forming a single-phase (if at all) solid-solution~\cite{adem2007,he2017formation,otto} in preference to competing phases such as intermetallics, precipitates, multiphase, and amorphous structures~\cite{sheng2011phase,ye2015,ye2016high}. It is implicitly evident from the existence of limited single-phase solid-solution HEAs.
\par
According to Hume-Rothery rules~\cite{hume1969structure}, a substitutional solid-solution forms if the constituent mixing elements have similar atomic sizes (radii difference $\le$ 15\%), electronegativities, valencies, and the same crystal structure. Generally, HEAs do not satisfy all of these rules \cite{troparevsky2015beyond}, and therefore, deviations from the ideal lattice structure are seen. These structural deviations, termed internal lattice distortions (ILDs), are the combined effect of the size mismatch, differences in constituent elemental crystal structures and their valencies, and bond-heterogeneities among mixing elements.
\par 
In this work, we demonstrate that the BCC structure of Nb$_x$(HfZrTi)$_y$ gets stabilized with increasing Nb-concentration, marking an HCP$\rightarrow$BCC transition. We show that the associated structural changes and variation in average number of valence electrons with the addition of Nb result in large fluctuations in ILDs. Through comparative analysis of Gibbs free energy of quaternary HEAs Nb$_x$(HfZrTi)$_y$ and Nb$_a$Mo$_b$W$_c$Ta$_d$, we find that the entropy stabilizes the former while the enthalpy of formation ensures the stability of the latter. 
\section{Computational Details}
We use special quasirandom structures (SQS)~\cite{zunger1990special,van2002alloy,van2013efficient} of HEAs to approximately model their chemical disorder. For each alloy, we generate SQS with 3$\times$3$\times$2 periodic supercell (36 atoms) of the conventional unit cells of BCC or HCP structures (see SI section I. for SQS details). We considered BCC and HCP host lattices of Nb$_x$(HfZrTi)$_y$ alloys and only the BCC lattice of Nb$_a$Mo$_b$W$_c$Ta$_d$. As the reference ideal solid solution for comparison, we chose SQS of completely miscible BCC Mo$_p$W$_q$ alloys for which the heat of mixing at any composition vanishes~\cite{murray1986binary, Colinet_1988}. Lattice parameters of SQS configurations were estimated using Vegard's law \cite{vegard_Ashcroft}, which were optimized through structural relaxation to an energy minimum.
\par 
We perform full structural relaxation of these model SQS within the density functional theory (DFT) methods incorporated in the Quantum ESPRESSO package~\cite{giannozzi2009quantum}. We used a generalized gradient approximation (GGA)~\cite{gga} and Perdew-Burke-Ernzerhof~\cite{pbe} functional of electronic exchange-correlation energy. We employ projector augmented wave potentials~\cite{paw} and represent the electronic wave functions and charge density with plane wave basis sets truncated at energy cutoffs of 60 Ry and 500 Ry respectively. Uniform meshes of $3\times3\times4$ and $3\times3\times3 \,\, k$-points were used in sampling integrations over Brillouin zones of BCC and HCP-based supercells respectively. Using Hellman-Feynman forces and Broyden Fletcher Goldfarb Shanno (BFGS) scheme, each alloy SQS is relaxed till the force components on each atom $\vec{F}$ becomes less than 10$^{-3}$ Ry/Bohr. Total energy was converged within 10$^{-8}$ Ry to achieve electronic self-consistency. Fermi-Dirac distribution with a width of $k_BT =$ 0.002 Ry is used for smoothening the discontinuity in occupation numbers of electronic states.
\par
We use the Debye model to estimate vibrational entropy. The Debye temperature ($\theta_D$) of each alloy SQS was extracted from their elastic moduli matrix obtained from the thermo\_pw package~\cite{dal}. $\theta_D$ is used as a single parameter within Debye model to estimate the vibrational entropy of each alloy.
\par 
To analyze the local structure of relaxed SQS lattices of HEAs, we investigate (a) the nearest-neighbor bond alignments of each atom through bond-orientational order parameters using a recently developed python library \emph{pyscal}~\cite{menon2019pyscal}, and (b) geometric features of the Voronoi cell constructed around each atom using \emph{Voro++} library~\cite{rycroft2009voro++} for Voronoi analysis. 
\section{Results and Discussion}
\begin{figure*}[htpb]
	\includegraphics[width=2.0\columnwidth]{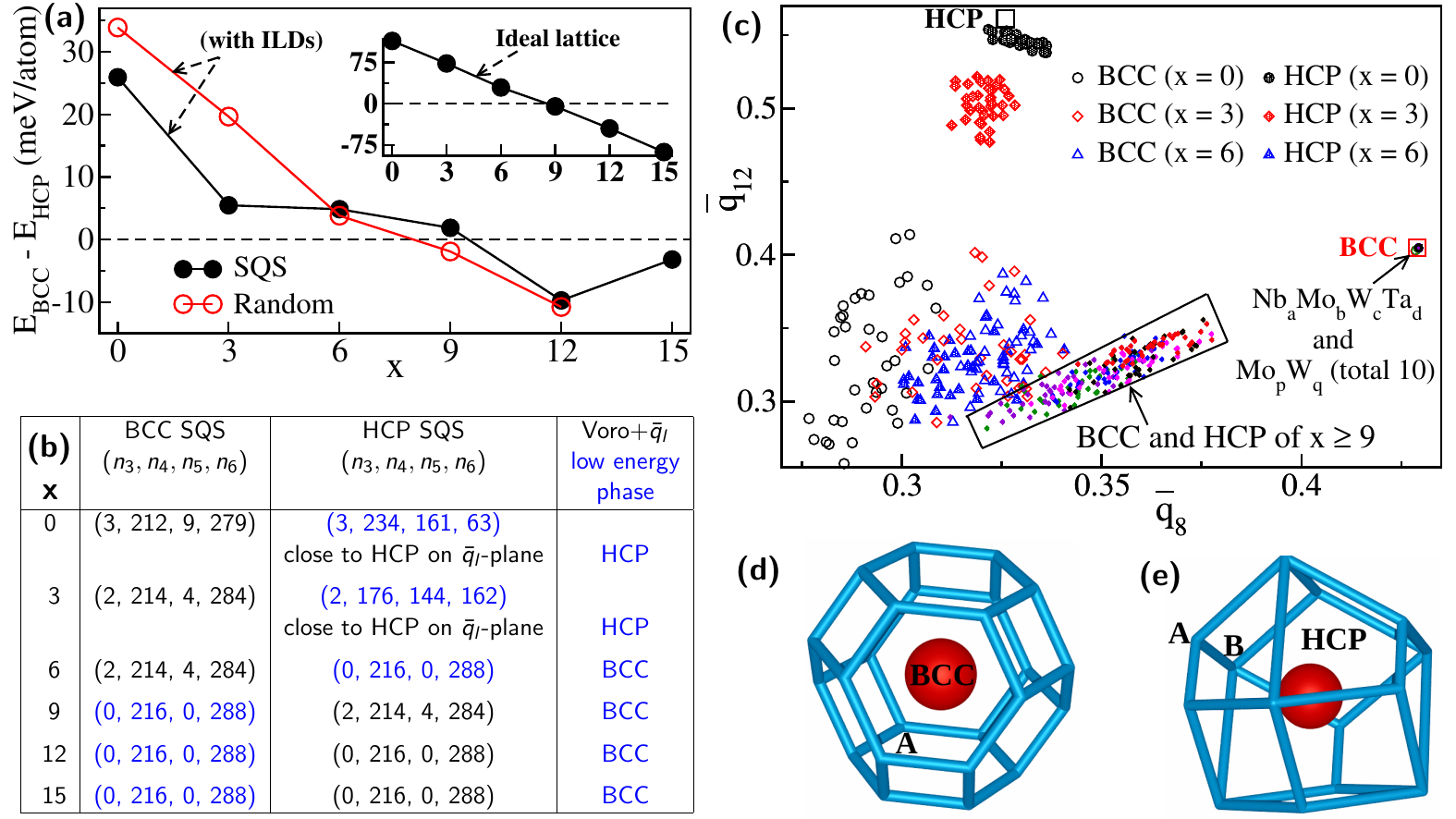}
	\caption{\label{fig1}Structural transition in Nb$_x$(HfZrTi)$_y$ with $x+3y=36$ and analysis with Voronoi tesselation (b) and bond orientational order parameters (c). In (a), relative energies of relaxed special quasirandom and randomized HCP and BCC structures of Nb$_x$(HfZrTi)$_y$ reveal stabilization of the BCC phase beyond 25\% Nb concentration (i.e; $x \geq 9$), corresponding to equiatomic composition. Unrelaxed SQS with ideal lattice (inset) also reveals the same. In (b), $n_k$ is the the number of $k$-sided Voronoi faces. Voronoi cell of an ideal BCC lattice point has squares and hexagons (d), while that of an HCP lattice point has only quadrilaterals (e). Voronoi vertex of order 3 (type-A) is stable and robust against internal lattice distortions (ILDs), while of the higher order (type-B) is unstable as new polygons appear there under distortion (see SI Fig. 1). From the Voronoi faces analysis (b) of energetically favorable SQS configurations as seen in (a), we find that a distorted BCC structure of Nb$_x$(HfZrTi)$_y$ is stabilized for $x\geq$ 6, corresponding to 16\% Nb-concentration rather than equiatomic concentration. (c) The plane of bond orientational parameters $(\bar{q}_{8}, \bar{q}_{12})$ facilitates identification of the lattice structure having low ILDs. It is clear that structures of Nb$_a$Mo$_b$W$_c$Ta$_d$ and Mo$_p$W$_q$ optimize to perfect BCC lattice with not much ILDs while Nb$_x$(HfZrTi)$_y$ exhibit severe ILDs as evident in a wide spread in their $\bar{q}_l$ values. The HCP SQS of low Nb-concentration ($x=$ 0 and 3) alloys optimize to points rather close to that of perfect HCP lattice on $(\bar{q}_{8}, \bar{q}_{12})$ plane. Thus we identify HCP$\rightarrow$BCC transition in Nb$_x$(HfZrTi)$_y$ at 16\% Nb-concentration.}
\end{figure*}

Our motivation for analysis of the relative stability of BCC and HCP structures of Nb$_x$(HfZrTi)$_y$ comes from the distinct crystal structures taken by its constituent elements at ambient conditions: Nb occurs in BCC structure, and Hf, Zr, and Ti occur in HCP structure. Secondly, Hf, Zr, and Ti undergo structural transformation to BCC structure at high temperature from their stable low-temperature HCP structure~\cite{fisher1964single}. Experimentally, the equiatomic NbHfZrTi alloy occurs in the BCC structure~\cite{ye2017dislocation,ye2018friction,lei2018enhanced}. Our calculations (Fig.~\ref{fig1}(a)) also support this as we find that at equiatomic and higher Nb-concentrations, SQS of Nb$_x$(HfZrTi)$_y$ of BCC lattice is more stable than that of HCP lattice. We find this structural preference in randomized structural models as well. With increasing Nb-concentration, Zhang et al.~\cite{zhang2018phase} report $\omega\,{(P6/mmm)} \rightarrow$ BCC transformation rather than HCP (P6$_3$/mmc) $\rightarrow$ BCC, which poses the question of the structure of low Nb-concentration Nb$_x$(HfZrTi)$_y$ alloys. To address this, we simulated a few SQS models of (HfZrTi)$_{12}$ considering HCP and $\omega$ lattices, and found that SQS on the HCP lattices were more stable than those on the $\omega$ ones (see SI Table I. and SI Fig. 3) by more than 15 meV/atom. Thus, we analyze here HCP to BCC transformation.
\par
In Fig.~\ref{fig1}(a), we report a structural transformation equiatomic composition of Nb$_x$(HfZrTi)$_y$ based on the relative DFT energies of SQS corresponding to HCP and BCC lattices. We find significant changes in the atomic positions of relaxed SQS from their ideal lattice structure; thus, ILDs help lower the energy. As these atomic displacements developed during relaxation severely distort the ideal lattice, we require local structural descriptors to ascertain the true phase of relaxed HEAs. At nonzero temperature, any crystalline material exhibits dynamic ILDs due to thermal vibrations. However, static ILDs are present in HEAs even at T = 0 K due to differences among their atomic constituents. In a recent study~\cite{role_ild}, root mean squared displacement of atoms in BCC structure of NbHfZrTi at T = 0 K was estimated at 9\% of its lattice parameter. Such large ILDs make identification of the lattice structure challenging. Experimentally, XRD gives the average crystalline structure with Bragg peaks broadened due to ILDs.
\subsection{Local structure and identification of the underlying lattice}
\paragraph{(a) Voronoi analysis} To examine the local structure at the atomic scale, we first use Voronoi decomposition involving the construction of a polyhedron around each lattice point (atomic site) known as Voronoi cell (or Wigner-Seitz cell in crystallography). Since the atomic radii of constituent atoms in HEAs are similar, we treated each atom as a point particle in Voronoi decomposition. The Voronoi cell of a BCC lattice point is a \emph{truncated octahedron} (see Fig.~\ref{fig1}(d)) that has 6 square and 8 hexagonal faces. The Voronoi cell of an HCP lattice point is a \emph{trapezo-rhombic dodecahedron} (see Fig.~\ref{fig1}(e)) that has 6 trapezium- and 6 rhombus-shaped faces. The order of a vertex in a graph is the number of edges incident into it. Each Voronoi vertex of a BCC lattice is of order 3, which is topologically stable (type-\textbf{A}). In contrast, some Voronoi vertices of the HCP lattice are of order 4, which are topologically unstable (type-\textbf{B}). With a slight perturbation, topologically unstable vertices modify the Voronoi cell characteristics by creating new faces~\cite{Troadec_1998}. We demonstrate this for supercell of BCC and HCP lattices by adding random displacements to the positions of their lattice points. While the Voronoi cells of the BCC lattice does not evolve to have any new polygonal faces, pentagonal and hexagonal faces appear in the Voronoi polyhedra of randomly perturbed HCP lattice originating at the unstable vertices of order 4 (see SI Fig. 2(b)).
\par 
We use this feature of geometric instability of Voronoi vertices to identify BCC and HCP lattices of minimum energy SQS of Nb$_x$(HfZrTi)$_y$ alloys (see Table contained in Fig.~\ref{fig1}), which exhibit significant ILDs. Voronoi's with only quadrilaterals and hexagonal faces reveal that alloys with significant Nb-concentration ($x\geq$ 6) optimize to a BCC lattice structure. We claim that at low Nb-concentration ($x =$ 0 and 3), SQS optimize to HCP-based structures where pentagonal and hexagonal Voronoi faces originate at type-\textbf{B} Voronoi vertices due to ILDs. 
\begin{figure*}[htp]
	\includegraphics[width=2.0\columnwidth]{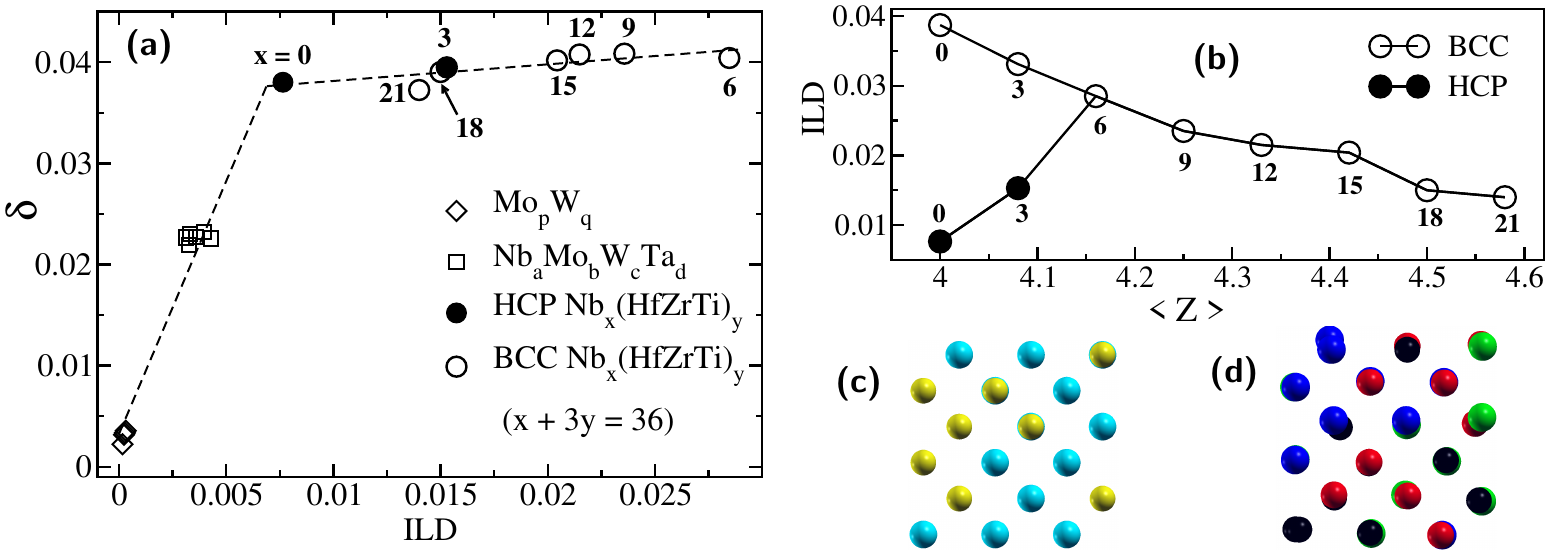}
	\caption{\label{fig2}Factors of ILDs: (a) In binary Mo$_p$W$_q$ (four different compositions), quaternary Nb$_a$Mo$_b$W$_c$Ta$_d$ (six different compositions), and Nb$_x$(HfZrTi)$_y$ (x is written near the symbols) alloys, internal lattice distortions (ILDs) are of the order of $10^{-4}, 10^{-3}, \textnormal{and}\, 10^{-2}$ respectively. The z-axis view of relaxed SQS of MoW (c) and NbHfZrTi (d) show significant presence of ILDs in the latter. Although atomic size mismatch ($\delta$) is a primary factor of ILDs, the significant variation in ILDs across different compositions of Nb$_x$(HfZrTi)$_y$ without any remarkable change in their $\delta$ values asks for another possible factor. In contrast to HEAs Nb$_a$Mo$_b$W$_c$Ta$_d$, the structural differences among the constituent elements of Nb$_x$(HfZrTi)$_y$ and associated HCP$\rightarrow$BCC transition at x = 6 where ILDs also peak (b) explains for the variation in their ILDs. For BCC lattice SQS, ILDs decrease with increase in average valency ($<Z>$) and that occurs when Nb-concentration (x) increases.}
\end{figure*}
\paragraph{(b) Bond-orientational order parameters} We chose another class of structural descriptors called bond-orientational order parameters to confirm our claim of structural transition. The local bond-orientational order parameters ($q_l$) proposed by Steinhardt \cite{steinhardt1983bond} capture the signatures of local structure. For each atom, $q_l$ is written in terms of spherical harmonics of $\theta_{ij}$ and $\phi_{ij}$ of orientational unit bond vectors joining neighboring sites $i$ and $j$:
\[q_{lm}(i) = \frac{1}{n(i)}\sum_{j=1}^{n(i)}Y_{lm}(\theta_{ij}, \phi_{ij})\]
\begin{equation}
	 q_l(i) = \sqrt{\frac{4\pi}{2l + 1}\sum_{m=-l}^{l}|q_{lm}(i)|^2},
\end{equation}
where $n(i)$ is the number of neighbor atoms around $i^{th}$ atom.

These local bond-orientational parameters are ultra-sensitive to the symmetry of the crystal and help identify simple phases such as BCC, FCC, and HCP. For different ideal crystals, $q_l$ values are distinct except for $q_l$ of odd $l$ (see SI Table II). ILDs or thermal noise disturb the crystal's local structure symmetry, and result in changes in $q_l$ values and complicate the identification of the average lattice structure. Lechner and Dellago~\cite{lechner2008accurate} showed that locally averaged $\bar{q}_l$s:

\[
	\bar{q}_{lm}(i) = \frac{1}{n(i)}\sum_{k=0}^{n(i)}q_{lm}(k) \]
\begin{equation}\bar{q}_l(i) = \sqrt{\frac{4\pi}{2l + 1}\sum_{m=-l}^{l}|\bar{q}_{lm}(i)|^2},
\end{equation}

work better in identifying the Bravais lattice of distorted crystals (see SI Fig. 4). For an ideal lattice, $\bar{q}_l$ coincides with $q_l$, and the separation between the two measures the loss of structural order. In resolution of the lattice structure, here, we use $(\bar{q}_8, \bar{q}_{12})$ plane in which ideal BCC and HCP lattices are well separated at points (0.429, 0.405) and (0.317, 0.565), respectively.
\par
In $(\bar{q}_8, \bar{q}_{12})$ plane (Fig.~\ref{fig1}(c)), relaxed SQS of Nb$_a$Mo$_b$W$_c$Ta$_d$ and Mo$_p$W$_q$ alloys always fall on to the point representing BCC lattice structure. In Nb$_x$(HfZrTi)$_y$ alloys, SQS of HCP lattices with low Nb-concentration ($x=$ 0 and 3) optimize to a structure that is close to the perfect HCP structure (here c/a = 1.60) with moderate distribution in $\bar{q}_l$, confirming that their relaxed lattices are distorted HCP structures. On the other hand, alloys with high Nb-concentration ($x\geq9$) represented with SQS of BCC or HCP lattices converge upon relaxation to the same domain in $(\bar{q}_8, \bar{q}_{12})$ plane (see also SI Fig. 5), and exhibit distorted BCC structures as revealed earlier here in the Voronoi analysis in Fig.~\ref{fig1}(b). 
\par Thus, we have explicitly shown with local structural analysis that Nb$_x$(HfZrTi)$_y$ alloys undergo a structural change from distorted HCP lattices (for $x=$ 0, and 3) to distorted BCC lattices (for $x \geq 6$) at 16\% Nb-concentration (since $100\times6/36\simeq16$). It is interesting that HCP lattice spontaneously transforms to BCC lattice through structural relaxation without having to cross any energy barrier. Secondly, this transformation has a signature in the electronic structures (see SI Fig. 8). At high Nb-concentration ($x\geq$ 9), $d$-orbitals of Nb dominate the electronic states near the Fermi energy, explaining how Nb is a BCC-stabilizer.
\subsection{ILDs peak at the HCP to BCC transition}
Wide distribution of $\bar{q}_l$ (Fig.~\ref{fig1}(c)) reflects on the presence of significant ILDs in HEAs Nb$_x$(HfZrTi)$_y$. For a quantitative measure of ILDs, we use~\cite{song2017local}
\begin{equation} 	
	\textnormal{ILD} = \frac{1}{N}\sum_{i=1}^N\sqrt{(x_i - x_i')^2 + (y_i - y_i')^2 + (z_i - z_i')^2}	
\end{equation}	
where $(x_i, y_i, z_i)$ and $(x_i', y_i', z_i')$ are reduced coordinates of unrelaxed sites (ideal, reference lattice points) and relaxed atomic positions of the $i^{th}$ atom, respectively and $N$ is the total number of atoms. We note that ILDs in binary Mo$_p$W$_q$, quaternary Nb$_a$Mo$_b$W$_c$Ta$_d$, and Nb$_x$(HfZrTi)$_y$ alloys are of order of $10^{-4}, 10^{-3}, \textnormal{and}\, 10^{-2}$, respectively (see Fig.~\ref{fig2}(a) and SI Table IV). In contrast to BCC (MoW)$_{18}$ (Fig.~\ref{fig2}(c)), a sideview of the relaxed SQS lattice of BCC (NbHfZrTi)$_{9}$ (Fig.~\ref{fig2}(d)) shows significant atomic perturbations from their ideal lattice sites. On increasing $x$ in Nb$_x$(HfZrTi)$_y$, ILDs of their energetically favorable SQS first increase, reach a maximum at $x= 6$ where HCP$\rightarrow$BCC transformation is marked, and then decrease subsequently (see Fig.~\ref{fig2}(b)). 
\par ILDs in HEAs originate from various factors such as atomic size mismatch, dissimilar crystal structures, and difference in valence electrons of their constituent elements. Multi-elemental mixing always suffer from atomic size mismatch ($\delta$) which is quantified as~\cite{adem2007}
\begin{equation}
	\delta^2 = \sum_{i=1}^{n}c_i(1 - r_i/\bar{r})^2 \hspace{0.25cm} \textnormal{with}\hspace{0.25cm}  \bar{r} = \sum_{i=1}^{n} c_ir_i,	
\end{equation}
where $n$ is the number of types of elements being mixed, $c_i$ and $r_i$ are the atomic concentration and radius of $i^{th}$ element, and $\bar{r}$ is the average radius.
$\delta$ is considered the primary factor relevant to ILDs~\cite{song2017local,wang2015}. 
\begin{figure*}[htp]
	\includegraphics[width=2.0\columnwidth]{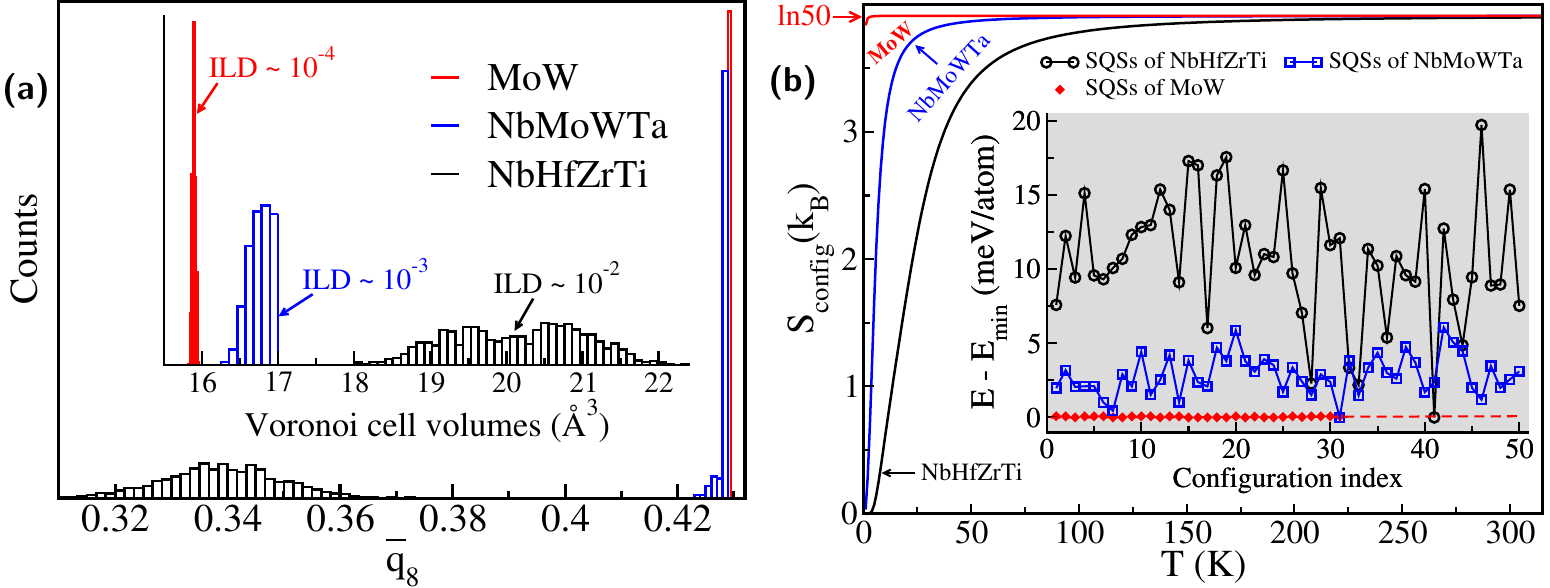}
	\caption{\label{fig3}Internal Lattice Distortions, energetics and configurational entropy. Histograms 
		(a) of bond orientational parameter $\bar{q}_8$ and Voronoi cell volumes (inset) obtained from the relaxed structures of 50 SQS configurations of each of the equiatomic BCC alloys. A sharp peak in $\bar{q}_8$ of MoW marks the ideal BCC structure, a slight deviation in NbMoWTa reveals relatively weak ILDs. In contrast, notably broad and shifted peak in $\bar{q}_8$ of NbHfZrTi reveals its severe ILDs, and distributions of Voronoi volumes confirms this trend in ILDs and the local structure. (b) Configurational entropy approaches the entropy of ideal mixing, (Boltzmann entropy $k_B\ln50$) at fairly low temperatures, though slower in NbHfZrTi due to larger fluctuations in the energy of their distinct SQS configurations (shown in inset) than of NbMoWTa and MoW.}	
\end{figure*}
Fig.~\ref{fig2}(a) depicts this for BCC Mo$_p$W$_q$, Nb$_a$Mo$_b$W$_c$Ta$_d$, and HCP (HfZrTi)$_{12}$ alloys as their ILDs linearly increase with $\delta$. However, taking Nb$_x$(HfZrTi)$_y$ alloys as an example, we demonstrate that rather than $\delta$, the elemental structural differences among their constituents and their average valence electrons (Fig.~\ref{fig2}(b)) strongly influence ILDs. An increase in the number of average valence electrons $<Z>$ within BCC Nb$_x$(HfZrTi)$_y$ --- achieved by the gradual rise of Nb-concentration --- lowers ILDs. This mechanism of control over ILDs by tuning $<Z>$ is consistent with a recent work reported only for BCC HEAs~\cite{role_ild}. Therefore, ILDs of Nb$_x$(HfZrTi)$_y$ alloys peak at the HCP$\rightarrow$BCC transition (see Fig.~\ref{fig2}(b)), and thus, ILDs in these alloys exhibit a dual effect of valency and crystal structures of constituent elements. The role of crystal structures of constituents on ILDs is further highlighted by another set of quaternary HEAs, Nb$_a$Mo$_b$W$_c$Ta$_d$ (composed of only BCC structural elements), which does not show noticeable variation in ILDs with compositions.
\subsection{ILDs and configurational entropy}
For a qualitative measure of ILDs, we generated and relaxed 50 distinct SQS configurations for each equiatomic BCC NbHfZrTi, NbMoWTa, and MoW alloys and analyzed them with histograms of their $\bar{q}_8$ values and Voronoi volumes (see Fig.~\ref{fig3}(a)). For NbHfZrTi, the distribution exhibits a very broad peak, while it is quite narrow for NbMoWTa, and MoW has a single sharp peak. The width of these peaks in distributions, similar to Bragg peak width in XRD, serves as a measure of ILDs. Thus, ILDs are negligible in MoW and notably significant in NbHfZrTi. Here, $\bar{q}_8$ of MoW corresponds to that of an ideal BCC structure ($\bar{q}_8 = q_8 = 0.429$).
\par To analyze the effects of ILDs on configurational energy of HEAs, we consider these equiatomic SQS configurations, each with a different chemical arrangement. We find that SQS configurations of HEAs span a range of energy while the configurations of a solid-solution (MoW) have almost the same energy (Fig.~\ref{fig3}(b) inset). Lower energy SQS configurations of HEAs are more favorable, and hence this energy fluctuation signifies a departure from the \emph{ideal} mixing condition that requires each configuration to have the same energy~\cite{he2017formation}. Clearly, HEAs with higher ILDs display large fluctuations in their configurational energies.
\par From the distribution of energy of these SQS configurations, we estimate configurational entropy. If energy of $i^{th}$ configuration be $E_i$ among chosen $\Omega_{\textnormal{config}}$ configurations, then at temperature T its probability will be 
\begin{equation}
	p_i = \frac{\exp(-\beta E_i)}{\sum_{1}^{\Omega_{\textnormal{config}}}\exp(-\beta E_i)}
\end{equation} where $\beta = 1/(k_BT)$. Configurational entropy is deduced as 
\begin{equation}
	S_{\textnormal{config}} = -k_B\sum_{i=1}^{\Omega_{\textnormal{config}}}p_i\ln p_i.
\end{equation} 
We note that $S_{\textnormal{config}}$ of NbMoWTa and NbHfZrTi rises with temperature (see Fig.~\ref{fig3}(b)) and saturates to the Boltzmann entropy $k_B\ln\Omega_{\textnormal{config}}$ (here, $k_B\ln50 = 3.912 k_B$ which corresponds to MoW). In the ideal mixing of $n$ types of elements, 
\begin{equation} S_{\textnormal{config}} = S_\textnormal{ideal}^\textnormal{mix} = -N_{\textnormal{atom}}k_B\sum_{i=1}^{n}c_i\ln c_i,
\end{equation}
where $c_i$ is the atomic concentration of $i^{th}$ element, and it becomes 
$S_{\textnormal{config}} = N_{\textnormal{atom}}k_B\ln n $
for the equiatomic case. The computed configurational entropy for a finite number of SQS configurations will be lower than their ideal mixing entropy (for equiatomic quaternary systems, $N_{\textnormal{atom}}k_B\ln n = 36 k_B\ln4 = 49.90 k_B >> k_B\ln50$), but it captures the thermal effect that the configurational entropy approaches the Boltzmann entropy as the temperature rises. It is noteworthy that before saturation, the increase in $S_{\textnormal{config}}$ with temperature is slower in HEAs exhibiting stronger ILDs (see Fig.~\ref{fig3}(b)). Thus, we demonstrate that it is reasonable to approximate the configurational entropy of HEAs as the ideal mixing entropy since each configuration becomes equiprobable even at a fairly low T (such as 300 K).
\begin{figure}[htp]
	\includegraphics[width=\columnwidth]{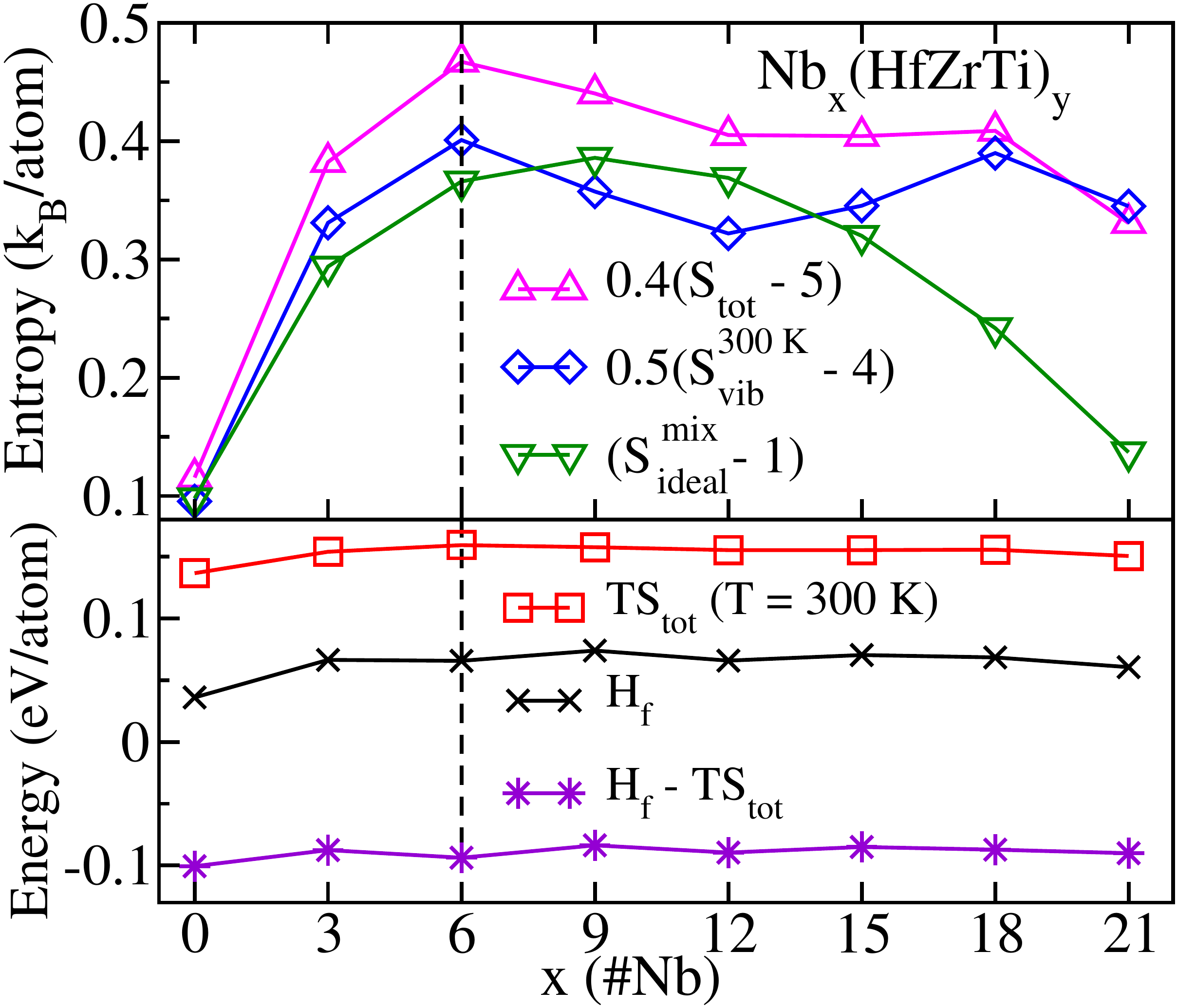}
	\caption{\label{fig4}Entropic stabilization of favorable phase Nb$_x$(HfZrTi)$_y$ where $x+3y = 36$. The total entropy comprised of Debye vibrational entropy and ideal mixing entropy reaches a maximum for BCC Nb$_6$(HfZrTi)$_{10}$ in contrast to configurational entropy (S$^\textnormal{mix}_\textnormal{ideal}$) that peaks at equiatomic (NbHfZrTi)$_9$ (top). Here, entropies have been scaled and shifted to facilitate comparison (see SI Table V for actual values). Entropic stabilization of 
	Nb$_x$(HfZrTi)$_y$ HEAs (bottom) is clear from the fact that their formation energies (H$_{{f}}$) are positive and entropic term, particularly with dominance of vibrational contribution (top), is needed to have favorable free energy of formation (bottom).}
\end{figure}
\subsection{Entropic stabilization} 
The postulate that high configurational entropy stabilizes the solid-solution phase of HEAs~\cite{yeh2004nanostructured} has been a topic of controversy~\cite{otto,more_entropy}. Other kinds such as vibrational, electronic, and magnetic entropies can also be important to the stability of HEAs, while major contributions come from vibrational and configurational ones~\cite{more_entropy}. In HEAs Nb$_x$(HfZrTi)$_y$, we find that vibrational entropy contributes more than the configurational one (see SI Table V) to the total entropy. We take the total entropy comprised of vibrational entropy estimated within Debye approximation and configurational entropy approximated as the entropy of ideal solid-solution mixing. In Fig.~\ref{fig4}(top), we show that the total entropy of non equiatomic Nb$_6$(HfZrTi)$_{10}$ is greater than that of equiatomic (NbHfZrTi)$_9$ peaking at $x$ = 6. It is interesting to note that $x$ = 6 marks the structural transformation from HCP to BCC, where ILDs and total entropy reach their maxima.
\par We estimate Gibbs free energy to assess the competition between formation energy (i.e., the heat of mixing) and entropy. For the formation energy ($H_f$) of alloys Nb$_x$(HfZrTi)$_y$, we subtract its concentration-weighted elemental energies in their most stable bulk crystalline phase from the energy of the alloy. The formation energy of Nb$_x$(HfZrTi)$_y$ is positive (sign of instability) while for Nb$_a$Mo$_b$W$_c$Ta$_d$ it is negative (see SI Table IV and V). It means that the former is unfavorable, and the latter is favorable energetically. However, inclusion of entropy makes Gibbs free energy of formation ($H_f - TS_\textnormal{tot}$) negative and stabilizes Nb$_x$(HfZrTi)$_y$ (see Fig.~\ref{fig4}(bottom)). For instance, $H_f$ of (NbHfZrTi)$_9$ is 0.074 eV/atom and at T = 300 K, alone $TS_{\textnormal{config}} = k_BT\ln4$ = 0.036 eV/atom is insufficient to stabilize and needed a major contribution from vibration as $TS_\textnormal{tot}$ is 0.158 eV/atom to achieve the overall stability (see SI Table V). Hence, with precise quantification, we reinforce the fundamental assumption that entropy stabilizes HEAs.
\section{Conclusions} In conclusion, we have shown that Nb$_x$(HfZrTi)$_y$ undergoes a structural transformation from HCP to BCC at 16\% Nb-concentration. Voronoi analysis and bond-orientational order parameters are tools to help identify the average lattice structure of HEAs exhibiting large ILDs and mark this transformation. The structural differences across constituent elements and their numbers of valence electrons are dominant factors that cause ILDs in addition to atomic size mismatch in BCC HEAs. At an HCP$\rightarrow$BCC transition, ILDs peak and maximize the total entropy. We showed that entropy stabilizes Nb$_x$(HfZrTi)$_y$, but with a larger share of the vibrational entropy than of the configurational entropy.

\section*{Acknowledgement} NK thanks the Materials Theory Group, JNCASR, for several fruitful discussions on this work in weekly group meetings. NK also acknowledges the Council of Scientific \& Industrial Research for the Ph.D. fellowship (Award No.- 09/733(0214)/2016-EMR-I). UVW acknowledges support from a JC Bose National fellowship of SERB-DST, Government of India.
\bibliographystyle{elsarticle-num}
\bibliography{arXiv.bib}

\begin{thebibliography}{10}
\expandafter\ifx\csname url\endcsname\relax
  \def\url#1{\texttt{#1}}\fi
\expandafter\ifx\csname urlprefix\endcsname\relax\def\urlprefix{URL }\fi
\expandafter\ifx\csname href\endcsname\relax
  \def\href#1#2{#2} \def\path#1{#1}\fi

\bibitem{yeh2004nanostructured}
J.-W. Yeh, S.-K. Chen, S.-J. Lin, J.-Y. Gan, T.-S. Chin, T.-T. Shun, C.-H.
  Tsau, S.-Y. Chang, Nanostructured high-entropy alloys with multiple principal
  elements: Novel alloy design concepts and outcomes, Adv. Eng. Mater. 6~(5)
  (2004) 299--303.
\newblock \href {http://dx.doi.org/10.1002/adem.200300567}
  {\path{doi:10.1002/adem.200300567}}.

\bibitem{cantor2004microstructural}
B.~Cantor, I.~Chang, P.~Knight, A.~Vincent, Microstructural development in
  equiatomic multicomponent alloys, Mater. Sci. Eng. A 375-377 (2004) 213--218.
\newblock \href {http://dx.doi.org/10.1016/j.msea.2003.10.257}
  {\path{doi:10.1016/j.msea.2003.10.257}}.

\bibitem{zhang2014}
Y.~Zhang, T.~T. Zuo, Z.~Tang, M.~C. Gao, K.~A. Dahmen, P.~K. Liaw, Z.~P. Lu,
  Microstructures and properties of high-entropy alloys, Progress in Materials
  Science 61 (2014) 1--93.
\newblock \href {http://dx.doi.org/10.1016/j.pmatsci.2013.10.001}
  {\path{doi:10.1016/j.pmatsci.2013.10.001}}.

\bibitem{miracle2017critical}
D.~Miracle, O.~Senkov, A critical review of high entropy alloys and related
  concepts, Acta Mater. 122 (2017) 448--511.
\newblock \href {http://dx.doi.org/10.1016/j.actamat.2016.08.081}
  {\path{doi:10.1016/j.actamat.2016.08.081}}.

\bibitem{grosse2018data}
S.~Gorsse, M.~Nguyen, O.~Senkov, D.~Miracle, Database on the mechanical
  properties of high entropy alloys and complex concentrated alloys, Data in
  Brief 21 (2018) 2664--2678.
\newblock \href {http://dx.doi.org/10.1016/j.dib.2018.11.111}
  {\path{doi:10.1016/j.dib.2018.11.111}}.

\bibitem{george2019high}
E.~P. George, D.~Raabe, R.~O. Ritchie, High-entropy alloys, Nat. Rev. Mater.
  4~(8) (2019) 515--534.
\newblock \href {http://dx.doi.org/10.1038/s41578-019-0121-4}
  {\path{doi:10.1038/s41578-019-0121-4}}.

\bibitem{murty2019high}
B.~S. Murty, J.-W. Yeh, S.~Ranganathan, P.~Bhattacharjee, High-entropy alloys,
  Elsevier, 2019.

\bibitem{li2019fcc}
Z.~Li, S.~Zhao, R.~O. Ritchie, M.~A. Meyers, Mechanical properties of
  high-entropy alloys with emphasis on face-centered cubic alloys, Progress in
  Materials Science 102 (2019) 296--345.
\newblock \href {http://dx.doi.org/10.1016/j.pmatsci.2018.12.003}
  {\path{doi:10.1016/j.pmatsci.2018.12.003}}.

\bibitem{pickering2016high}
E.~J. Pickering, N.~G. Jones, High-entropy alloys: a critical assessment of
  their founding principles and future prospects, Int. Mater. Rev. 61~(3)
  (2016) 183--202.
\newblock \href {http://dx.doi.org/10.1080/09506608.2016.1180020}
  {\path{doi:10.1080/09506608.2016.1180020}}.

\bibitem{adem2007}
Y.~Zhang, Y.~Zhou, J.~Lin, G.~Chen, P.~Liaw, Solid-solution phase formation
  rules for multi-component alloys, Advanced Engineering Materials 10~(6)
  (2008) 534--538.
\newblock \href {http://dx.doi.org/10.1002/adem.200700240}
  {\path{doi:10.1002/adem.200700240}}.

\bibitem{he2017formation}
Q.~He, Y.~Ye, Y.~Yang, Formation of random solid solution in multicomponent
  alloys: from {Hume-Rothery} rules to entropic stabilization, J. Phase
  Equilibria Diffus. 38~(4) (2017) 416--425.
\newblock \href {http://dx.doi.org/10.1007/s11669-017-0560-9}
  {\path{doi:10.1007/s11669-017-0560-9}}.

\bibitem{otto}
F.~Otto, Y.~Yang, H.~Bei, E.~George, Relative effects of enthalpy and entropy
  on the phase stability of equiatomic high-entropy alloys, Acta Mater. 61~(7)
  (2013) 2628--2638.
\newblock \href {http://dx.doi.org/10.1016/j.actamat.2013.01.042}
  {\path{doi:10.1016/j.actamat.2013.01.042}}.

\bibitem{sheng2011phase}
S.~Guo, C.~T. Liu, Phase stability in high entropy alloys: Formation of
  solid-solution phase or amorphous phase, Prog. Nat. Sci.: Mater. Int. 21~(6)
  (2011) 433--446.
\newblock \href {http://dx.doi.org/10.1016/S1002-0071(12)60080-X}
  {\path{doi:10.1016/S1002-0071(12)60080-X}}.

\bibitem{ye2015}
Y.~Ye, Q.~Wang, J.~Lu, C.~Liu, Y.~Yang, Design of high entropy alloys: A
  single-parameter thermodynamic rule, Scr. Mater. 104 (2015) 53--55.
\newblock \href {http://dx.doi.org/10.1016/j.scriptamat.2015.03.023}
  {\path{doi:10.1016/j.scriptamat.2015.03.023}}.

\bibitem{ye2016high}
Y.~F. Ye, Q.~Wang, J.~Lu, C.~T. Liu, Y.~Yang, High-entropy alloy: challenges
  and prospects, Mater. Today 19~(6) (2016) 349--362.
\newblock \href {http://dx.doi.org/10.1016/j.mattod.2015.11.026}
  {\path{doi:10.1016/j.mattod.2015.11.026}}.

\bibitem{hume1969structure}
W.~Hume-Rothery, R.~Smallman, C.~Haworth, The structure of metals and alloys,
  The Institute of Metals, London (1969).

\bibitem{troparevsky2015beyond}
M.~C. Troparevsky, J.~R. Morris, M.~Daene, Y.~Wang, A.~R. Lupini, G.~M. Stocks,
  Beyond atomic sizes and hume-rothery rules: understanding and predicting
  high-entropy alloys, Jom 67~(10) (2015) 2350--2363.
\newblock \href {http://dx.doi.org/10.1007/s11837-015-1594-2}
  {\path{doi:10.1007/s11837-015-1594-2}}.

\bibitem{zunger1990special}
A.~Zunger, S.-H. Wei, L.~G. Ferreira, J.~E. Bernard, Special quasirandom
  structures, Phys. Rev. Lett. 65 (1990) 353--356.
\newblock \href {http://dx.doi.org/10.1103/PhysRevLett.65.353}
  {\path{doi:10.1103/PhysRevLett.65.353}}.

\bibitem{van2002alloy}
A.~{van de Walle}, M.~Asta, G.~Ceder, The alloy theoretic automated toolkit: A
  user guide, Calphad 26~(4) (2002) 539--553.
\newblock \href {http://dx.doi.org/10.1016/S0364-5916(02)80006-2}
  {\path{doi:10.1016/S0364-5916(02)80006-2}}.

\bibitem{van2013efficient}
A.~{van de Walle}, P.~Tiwary, M.~{de Jong}, D.~Olmsted, M.~Asta, A.~Dick,
  D.~Shin, Y.~Wang, L.-Q. Chen, Z.-K. Liu, Efficient stochastic generation of
  special quasirandom structures, Calphad 42 (2013) 13--18.
\newblock \href {http://dx.doi.org/10.1016/j.calphad.2013.06.006}
  {\path{doi:10.1016/j.calphad.2013.06.006}}.

\bibitem{murray1986binary}
J.~L. Murray, L.~H. Bennett, H.~Baker, Binary alloy phase diagrams, Vol.~2, ASM
  International (OH), 1986.

\bibitem{Colinet_1988}
C.~Colinet, A.~Bessoud, A.~Pasturel, Theoretical determinations of
  thermodynamic data and phase diagrams of {BCC} binary transition-metal
  alloys, J. Phys. F: Met. Phys. 18~(5) (1988) 903--921.
\newblock \href {http://dx.doi.org/10.1088/0305-4608/18/5/010}
  {\path{doi:10.1088/0305-4608/18/5/010}}.

\bibitem{vegard_Ashcroft}
A.~R. Denton, N.~W. Ashcroft, Vegard's law, Phys. Rev. A 43 (1991) 3161--3164.
\newblock \href {http://dx.doi.org/10.1103/PhysRevA.43.3161}
  {\path{doi:10.1103/PhysRevA.43.3161}}.

\bibitem{giannozzi2009quantum}
P.~Giannozzi, S.~Baroni, N.~Bonini, M.~Calandra, R.~Car, C.~Cavazzoni,
  D.~Ceresoli, G.~L. Chiarotti, M.~Cococcioni, I.~Dabo, et~al., Quantum
  espresso: a modular and open-source software project for quantum simulations
  of materials, J. Phys.: Condens. Matter 21~(39) (2009) 395502.
\newblock \href {http://dx.doi.org/10.1088/0953-8984/21/39/395502}
  {\path{doi:10.1088/0953-8984/21/39/395502}}.

\bibitem{gga}
J.~P. Perdew, J.~A. Chevary, S.~H. Vosko, K.~A. Jackson, M.~R. Pederson, D.~J.
  Singh, C.~Fiolhais, Atoms, molecules, solids, and surfaces: Applications of
  the generalized gradient approximation for exchange and correlation, Phys.
  Rev. B 46 (1992) 6671--6687.
\newblock \href {http://dx.doi.org/10.1103/PhysRevB.46.6671}
  {\path{doi:10.1103/PhysRevB.46.6671}}.

\bibitem{pbe}
J.~P. Perdew, K.~Burke, M.~Ernzerhof, Generalized gradient approximation made
  simple, Phys. Rev. Lett. 77 (1996) 3865--3868.
\newblock \href {http://dx.doi.org/10.1103/PhysRevLett.77.3865}
  {\path{doi:10.1103/PhysRevLett.77.3865}}.

\bibitem{paw}
P.~E. Bl\"ochl, Projector augmented-wave method, Phys. Rev. B 50 (1994)
  17953--17979.
\newblock \href {http://dx.doi.org/10.1103/PhysRevB.50.17953}
  {\path{doi:10.1103/PhysRevB.50.17953}}.

\bibitem{dal}
\url{https://github.com/dalcorso/thermo_pw}.

\bibitem{menon2019pyscal}
S.~Menon, G.~D. Leines, J.~Rogal, pyscal: A python module for structural
  analysis of atomic environments, J. Open Source Softw. 4~(43) (2019) 1824.
\newblock \href {http://dx.doi.org/10.21105/joss.01824}
  {\path{doi:10.21105/joss.01824}}.

\bibitem{rycroft2009voro++}
C.~Rycroft, Voro++: A three-dimensional voronoi cell library in c++, Tech.
  rep., Lawrence Berkeley National Lab.(LBNL), Berkeley, CA (United States)
  (2009).
\newblock \href {http://dx.doi.org/10.2172/946741} {\path{doi:10.2172/946741}}.

\bibitem{fisher1964single}
E.~S. Fisher, C.~J. Renken, Single-crystal elastic moduli and the hcp
  \ensuremath{\rightarrow} bcc transformation in {Ti, Zr, and Hf}, Phys. Rev.
  135 (1964) A482--A494.
\newblock \href {http://dx.doi.org/10.1103/PhysRev.135.A482}
  {\path{doi:10.1103/PhysRev.135.A482}}.

\bibitem{ye2017dislocation}
Y.~X. Ye, Z.~P. Lu, T.~G. Nieh, Dislocation nucleation during nanoindentation
  in a body-centered cubic {TiZrHfNb} high-entropy alloy, Scr. Mater. 130
  (2017) 64--68.
\newblock \href {http://dx.doi.org/10.1016/j.scriptamat.2016.11.019}
  {\path{doi:10.1016/j.scriptamat.2016.11.019}}.

\bibitem{ye2018friction}
Y.~X. Ye, C.~Z. Liu, H.~Wang, T.~G. Nieh, Friction and wear behavior of a
  single-phase equiatomic {TiZrHfNb} high-entropy alloy studied using a
  nanoscratch technique, Acta Mater. 147 (2018) 78--89.
\newblock \href {http://dx.doi.org/10.1016/j.actamat.2018.01.014}
  {\path{doi:10.1016/j.actamat.2018.01.014}}.

\bibitem{lei2018enhanced}
Z.~Lei, X.~Liu, Y.~Wu, H.~Wang, S.~Jiang, S.~Wang, X.~Hui, Y.~Wu, B.~Gault,
  P.~Kontis, et~al., Enhanced strength and ductility in a high-entropy alloy
  via ordered oxygen complexes, Nature 563~(7732) (2018) 546--550.
\newblock \href {http://dx.doi.org/10.1038/s41586-018-0685-y}
  {\path{doi:10.1038/s41586-018-0685-y}}.

\bibitem{zhang2018phase}
L.~Zhang, H.~Fu, S.~Ge, Z.~Zhu, H.~Li, H.~Zhang, A.~Wang, H.~Zhang, Phase
  transformations in body-centered cubic {NbxHfZrTi} high-entropy alloys,
  Mater. Charact. 142 (2018) 443--448.
\newblock \href {http://dx.doi.org/10.1016/j.matchar.2018.06.012}
  {\path{doi:10.1016/j.matchar.2018.06.012}}.

\bibitem{role_ild}
G.~D. Samolyuk, Y.~N. Osetsky, G.~M. Stocks, J.~R. Morris, Role of static
  displacements in stabilizing body centered cubic high entropy alloys, Phys.
  Rev. Lett. 126 (2021) 025501.
\newblock \href {http://dx.doi.org/10.1103/PhysRevLett.126.025501}
  {\path{doi:10.1103/PhysRevLett.126.025501}}.

\bibitem{Troadec_1998}
J.~P. Troadec, A.~Gervois, L.~Oger, Statistics of voronoi cells of slightly
  perturbed face-centered cubic and hexagonal close-packed lattices, Europhys.
  Lett. 42~(2) (1998) 167--172.
\newblock \href {http://dx.doi.org/10.1209/epl/i1998-00224-x}
  {\path{doi:10.1209/epl/i1998-00224-x}}.

\bibitem{steinhardt1983bond}
P.~J. Steinhardt, D.~R. Nelson, M.~Ronchetti, Bond-orientational order in
  liquids and glasses, Phys. Rev. B 28 (1983) 784--805.
\newblock \href {http://dx.doi.org/10.1103/PhysRevB.28.784}
  {\path{doi:10.1103/PhysRevB.28.784}}.

\bibitem{lechner2008accurate}
W.~Lechner, C.~Dellago, Accurate determination of crystal structures based on
  averaged local bond order parameters, J. Chem. Phys. 129~(11) (2008) 114707.
\newblock \href {http://dx.doi.org/10.1063/1.2977970}
  {\path{doi:10.1063/1.2977970}}.

\bibitem{song2017local}
H.~Song, F.~Tian, Q.-M. Hu, L.~Vitos, Y.~Wang, J.~Shen, N.~Chen, Local lattice
  distortion in high-entropy alloys, Phys. Rev. Mater. 1 (2017) 023404.
\newblock \href {http://dx.doi.org/10.1103/PhysRevMaterials.1.023404}
  {\path{doi:10.1103/PhysRevMaterials.1.023404}}.

\bibitem{wang2015}
Z.~Wang, W.~Qiu, Y.~Yang, C.~Liu, Atomic-size and lattice-distortion effects in
  newly developed high-entropy alloys with multiple principal elements,
  Intermetallics 64 (2015) 63--69.
\newblock \href {http://dx.doi.org/10.1016/j.intermet.2015.04.014}
  {\path{doi:10.1016/j.intermet.2015.04.014}}.

\bibitem{more_entropy}
D.~Ma, B.~Grabowski, F.~K{\"o}rmann, J.~Neugebauer, D.~Raabe, Ab initio
  thermodynamics of the {CoCrFeMnNi} high entropy alloy: Importance of entropy
  contributions beyond the configurational one, Acta Mater. 100 (2015) 90--97.
\newblock \href {http://dx.doi.org/10.1016/j.actamat.2015.08.050}
  {\path{doi:10.1016/j.actamat.2015.08.050}}.

\end{thebibliography}
\end{document}